# Poverty and Perceptions of Electoral Integrity in the U.S.

*Economics & Politics, forthcoming (2025)*

**Douglas Cumming**
School of Business
Stevens Institute of Technology
525 River St.
Hoboken, New Jersey, 07030
Email: dcumming@stevens.com

**Sofia Johan**
College of Business,
Florida Atlantic University
777 Glades Road
Boca Raton, Florida 33431
Email: sofia.a.johan@gmail.com

**Ikenna Uzuegbunam**
Department of Management
School of Business
Howard University
2600 6th Street NW, Washington D.C. 20059
Email : ikennauzuegbunam@gmail.com

# Poverty and Perceptions of Electoral Integrity in the U.S.

**ABSTRACT**

We propose two opposing forces that impact the relation between electoral integrity and poverty. On the one hand, it is more costly to provide electoral integrity in states where there is more poverty due to transaction costs and opportunity costs. On the other hand, extreme levels of poverty attract media scrutiny and greater external monitoring of electoral integrity, giving rise to more demand for electoral integrity. Taken together, we expect electoral integrity to be a U-shaped function of poverty. We also hypothesize that electoral integrity will vary depending on the strength of state electoral laws. Expert-level survey data on electoral integrity from the 2016 U.S. Presidential election and the 2018 U.S. congressional election, in combination with U.S. state-level data on poverty are strongly consistent with these predictions.

# Poverty and Perceptions of Electoral Integrity in the U.S.

## 1. INTRODUCTION

Integrity, fairness, and quality of the electoral process are core features of formal institutions in democratic societies (Akoz & Arbatli, 2016; Ashworth, Bueno de Mesquita, & Friedenberg, 2017; Chaves, Fergusson, & Robinson, 2015; Henninger, Meredith & Morse, 2021; Hill, 2015; Kam & Mikos, 2007; Kratou & Laakso, 2022; Linebarger & Salehyan, 2020; Niven, 2022). Electoral integrity refers to "agreed international principles, values, and standards of elections, applying universally to all countries worldwide throughout the electoral cycle, including during the pre-electoral period, the campaign, on polling day, and its aftermath." (Norris, 2013a: p. 579). Electoral integrity is synonymous with 'free and fair elections' and 'electoral quality' and is prerequisite for sustainable development and poverty alleviation (Uberti & Jackson, 2020). To the extent that a democracy has accumulated a consistent record of multiple free and fair election cycles, it will be more likely to be perceived as having greater electoral integrity and more developed institutions. As a result of this assumption, many scholars have tended to focus attention on electoral integrity problems in younger and unsteady democracies, where there are high levels of electoral mistrust (see Cantu, 2014; Cantu & Garcia-Ponce, 2015; Jensen & Justesen, 2014; Kratou & Laakso, 2022). That notwithstanding, advanced democracies such as the U.S. are still susceptible to electoral malpractices (Jimenez, Hidalgo & Klimek, 2017; Korte, 2019; Norris, 2013b). In fact, questions related to declining electoral integrity in the U.S. have become a major policy issue according to bipartisan think tanks (e.g., Freedom House, 2020). Weichelt (2022) argues that Americans confidence in the US electoral system "is arguably at an all-time low" (p. 4). Regardless of whether these concerns are warranted, negative perceptions about electoral integrity in a society can reduce public confidence in the legitimacy of the electoral process (Karp, Nai, & Norris, 2018; Mongrain, 2023). It is also likely to have an impact on the development trajectory of affected economies.

We build on prior studies by both conceptually and empirically examining for the first time the intersection between poverty and electoral integrity. We develop novel predictions using straightforward economic concepts pertaining to the demand and supply of electoral integrity. We conjecture a non-linear relation between the variables based on asymmetric responses to moderate versus extreme poverty.

We relate electoral integrity data to state-level poverty rates across the US. This subnational focus on the electoral process within a country permits us to hold several relevant national level factors constant in the empirical setup (e.g., nature and length of democracy, cultural and environmental factors; Gupta & Panagariya, 2014). In addition, we account for other factors that vary across states including but not limited Republican versus Democratic state leadership and state electoral laws. The data show strong evidence that electoral integrity is a U-shaped function of poverty. Electoral integrity is indirectly facilitated by the fairness of state electoral laws. The findings are robust to the inclusion of the winner effect (or loser effect) in the modeling of expert opinions on electoral integrity (Sinclair, Smith & Tucker, 2018). That is, the results remain regardless of whether the opinions were derived from the subset of experts that supported the winning or losing candidate.

The findings of this research contribute to better understanding of the interaction between macro-level proxies of economic development, and individual-level perceptions of the legitimacy of the democratic process. Specifically, by leveraging the variance across US states in terms of poverty and electoral integrity, we highlight how certain states in the middle of the US poverty distribution may be shortchanged by supply and demand for electoral integrity. Moreover, by revealing the mediating effect of electoral laws in the analysis of poverty and electoral integrity, this study suggests some promising avenues for future electoral policy.

## 2. HYPOTHESES

Prior research indicates that there is a link between economic conditions of a geography and electoral process and outcomes (Hickey, 2009). Despite this well-established connection in the prior literature, the specific linkage between economic poverty and electoral integrity remains unclear. Why would poverty affect electoral integrity? A greater supply of electoral integrity costs more. But the costs of providing the same level of electoral integrity are not uniform across states. There are two reasons why electoral integrity costs more in poorer states with moderate or high rates of poverty than in richer states with low poverty rates. First, there are direct transaction costs associated with engaging with the electorate in poorer states. Poorer states require more institutional improvements, not just in their electoral systems, but also in their complementary systems to achieve comparable electoral integrity levels as richer states. According to Linz and Stephan (1996) cited in Norris et al., (2014, p.792) 'many other essential political rights and civil liberties need to be established for democratic consolidation to occur, notably the development of flanking institutions, including effective legislatures and independent courts with the capacity to check the power of the executive and maintain rule of law, an independent and pluralistic mass media, and a flourishing civic society'. Thus, the perceived transaction costs in poorer states will seem larger than richer states that already have more highly developed flanking institutions. Moreover, high poverty regions are often the product of being less socially minded; and in less socially minded groups, it is harder and costlier to ensure and provide integrity.

The second reason involves opportunity costs. Due to finite campaign budgets and state budgets for electoral processes, we expect that electoral integrity costs have the potential to compete with spending on other immediate economic priorities which are more pronounced in poorer states. Research evidence shows that poverty forces people to be more short-sighted in their budgeting (Haushofer & Fehr, 2014). From a state-level perspective, governments in poorer states will have more constraints on their budgets, which might lead them to spend less in improving their electoral infrastructure, policies, and systems (Daily Reporter, 2020).

Taken together, the supply of electoral integrity comes at a cost, which leads to an upward sloping supply curve for electoral integrity as depicted in Figure 1. But the transaction and opportunity costs of providing electoral integrity flatten the supply curve of electoral integrity in Figure 1 in states with moderate or high levels of poverty. The marginal costs of providing more electoral integrity are more pronounced depending on the lack of infrastructure in a state.

----Insert Figure 1 about here----

The demand for electoral integrity slopes downwards in Figure 1. Citizens would like more electoral integrity, but at a declining rate when they see it comes at a cost towards their tax dollars. External monitoring and media can exogenously shift the demand for electoral integrity (Grömping, 2019; Norris, Cameron, & Wynter, 2019), thereby leading to an increase in its provision. External election monitoring and media coverage are costly and limited and cannot be universally provided to all states. The returns to election monitoring and media coverage are more pronounced from engaging in regions that are more salient. Regions with most pronounced poverty rates and facing extreme needs tend to attract media attention and external aid, while the same provision is not uniformly provided to other less extreme regions (Terleckyj, 1970; Page & Pande, 2018; Vossen, van Gorp, & Schulpen, 2018).

There is an important difference in extreme poverty and moderate poverty in respect of the demand for electoral integrity. Extreme poverty shifts reliance on states internal resources to more access to external resources. Extreme poverty gathers more media attention to the electoral processes and systems, thereby helping increase external electoral resources and scrutiny in the form of electoral observers, more media attention, and lobbying resources for improved election laws. Extreme poverty shifts the demand curve for electoral integrity to the right, unlike moderate poverty. In net, therefore, we expect that poverty has a quadratic impact on electoral integrity, with electoral integrity first declining at modest rates, and then improving with extreme poverty rates.

**Hypothesis 1:** *Poverty has a quadratic impact on electoral integrity, with electoral integrity first declining at modest rates, and then improving with extreme poverty rates.*

Electoral monitoring is ineffective in giving rise to improved electoral integrity without strong election laws (Yukawa, 2018). Electoral laws ensure that in areas with high poverty rates, there are minimum standards that need to be upheld. Examples of electoral laws improvements that are aimed at strengthening electoral integrity include campaign finance and affirmative action laws (Uberti & Jackson, 2020). Improvements in electoral laws facilitate a culture that accepts the importance of electoral fairness. Electoral laws that are designed to be non-partisan improve electoral outcomes, while electoral laws that favor incumbents reduce electoral integrity (Frank & Coma, 2017). Electoral laws thus have the potential to mediate the impact of poverty on electoral integrity through ensuring fair standards despite the greater costs associated with providing electoral in poverty-stricken regions.

**Hypothesis 2:** *The fairness of electoral laws will mediate the impact of poverty on electoral integrity.*

# 3. DATA AND EMPIRICAL MEASURES

## 3.1 Data

The empirical context is the US presidential election on November 8, 2016, and the US congressional elections on November 6, 2018. The 2016 presidential election is particularly useful for studying electoral integrity in the United States because of the heightened concerns about electoral integrity from both the winning and losing party in the election. Nevertheless, we included the 2018 election data in the analysis to ensure that our claims are consistent over time. Focusing our analysis within the United States allows us to automatically control for other relevant factors that might vary across countries (Gupta & Panagariya, 2014). The data for this study comes from the Perceptions of Electoral Integrity US 2016 subnational study and US 2018 subnational study datasets (i.e., PEI US dataset), which is based on a survey of experts administered by the Electoral Integrity Project (Norris, Nai, & Gromping, 2017). According to Norris, Nai and Gromping (2017), an expert in the context of this study is "a political scientist (or social scientist in a related discipline such as law, sociology, economics, or anthropology) who has published on (or who has other demonstrated knowledge of) the electoral process in a particular country" (p.3).

The initial Perceptions of Electoral Integrity (PEI, 2016) survey was administered online, and experts were contacted two weeks after the US presidential election on November 8, 2016. Survey administrators subsequently sent three reminders on November 29, December 6, and December 12, 2016, yielding 726 responses across 50 U.S. states and the District of Columbia, corresponding to an overall response rate of 18.9% (Norris, Nai & Gromping, 2017). Similarly, the respondents to the (PEI, 2018) survey were first asked to respond to the survey four weeks after the congressional election, on December 4, 2018. Like the 2016 survey, three reminders were sent on December 11, December 18 and December 25, 2018, respectively. The 2018 survey yielded responses from 574 experts. Thus, the total number of respondents from the 2016 and 2018 surveys is 1300 experts. The current study leveraged both the expert-level data on 1300 respondents, as well as aggregated state-level data on 50 states and the District of Columbia. The aggregated state-level data on 50 states and D.C for two time periods, 2016 and 2018 produced a panel dataset. We merged the PEI US dataset with state-level data from official US government sources such as the US Census Bureau, Bureau of Labor Statistics, Department of Commerce, Department of Labor and Social Security Administration.

**3.2 Dependent variable**

There are three alternate measures of state-level *electoral integrity* available in the PEI US dataset. The first measure is an overall measure of the *electoral integrity rating* in a state. This measure is drawn from a survey question "*Overall, how would you rate the integrity of the US presidential election as it was conducted in [STATE] on a scale from 1 (very poor) to 10 (very good)*?" (Norris, Nai & Gromping, 2017). The second measure is the *PEI index of electoral integrity,* which is based on a sum of answers to 49 substantive variables in the survey, standardized to a 100-point scale and is imputed for situations where an expert did not answer a question (Norris, Nai, & Gromping, 2017; details are available in the codebook excerpts in appendices). This alternate measure provides an overall summary evaluation of expert perceptions that an election is consistent with international standards and global norms. The third measure of electoral integrity, which is also the mediating variable in the framework, is a measure of the fairness of *electoral laws index* in a state. We note that while the *electoral laws index* is a component of the *PEI index of electoral integrity* described above, it is a separate survey measure from the primary *electoral integrity rating* measure. This *electoral laws index* measure coded and standardized to a 100-point scale, captures responses to three questions regarding the extent to which electoral laws reflected fairness to smaller parties, favoritism to incumbent (governing) parties, and restrictions on citizens' rights in a state. All three measures of electoral integrity are based on responses related to the 2016 U.S. Presidential election and 2018 midterm election.

**3.3 Explanatory variables**

We measured state-level *poverty* as the percentage of poor in a given state based on U.S. Census Bureau definition of poverty thresholds (Osberg, 2000), in the preceding year, 2015 and 2017 respectively.

**3.4 Control variables**

To account for the influence of other individual level factors on experts' assessment of electoral integrity, we controlled for salient individual factors including gender, whether the respondent is a U.S. citizen, whether the respondent was born in the state, or have lived in a state, and whether the respondent supported the winning or the losing candidate. We also controlled for perceptions of the country's electoral integrity and whether the election was rigged.

     Our measure of party affiliation is a dummy variable, which is coded as 1 if the governor of the state is a Democrat and coded as 0 if the governor is Republican. Modern Democratic Party ideology emphasizes values that favor equal rights, while Republican party ideology emphasizes limited government intervention and self-reliance (Goren, 2004). State politics, especially in instances of a divided government, therefore can have a direct effect on the level of electoral integrity and electoral outcomes (Engstrom & Kernell, 2005). Finally, we controlled for the political heterogeneity in the state legislature by include a variable that captures the proportion of democrats in the state house and the state senate. This average measure across both state house and state senate. Our state-level control measures were also lagged one year, like the explanatory variable.

----Insert Tables 1 and 2 about here----

# 4. RESULTS

## 4.1 Summary statistics and univariate analyses

Table 1 shows the summary statistics and bivariate correlations for the key variables of this study. We note that all three measures of electoral integrity are significantly correlated ($r = 0.721, 0.515, 0.649, p<0.01$). Furthermore, the correlation matrix indicates that the main effect of poverty has a negative and statistically significant correlation with these measures of electoral integrity ($p<0.01$). We also note that the Governor affiliation (Democrat =1), and State house/senate measures are positively and significantly correlated with the three measures of electoral integrity ($p<0.01$).

## 4.2 Expert-level multivariate analyses

Our theoretical reasoning suggests that the effect of poverty on *electoral integrity* is attributed to the sensitivity over time on how *poverty* may undermine *electoral integrity*. We first conducted multivariate analysis of the multi-level (expert-level and state-level) data using ordinary least squares regression with standard errors clustered by states (see Huber 2012 for similar methodology). The results of our main multivariate analyses are reported in Tables 2 – 3. Poverty is negative and statistically significant in all models in Tables 2 – 3 (i.e., Models 1 – 11, $p<0.01$). Economically, the effect of poverty is equally significant. For example, one standard deviation increase in state poverty rate (i.e., a 2.44% increase) is associated with 3.82% decrease in *state rating in electoral integrity* (Model 2), a 3.32% decrease in *PEI-Index for electoral integrity* (Model 6), and a 7.18% decrease in electoral law index (Model 9). Furthermore, the squared term of poverty is positive and statistically significant in Table 2 models ($p<0.05$).

----Insert Tables 2 and 3 about here----

Figures 2a, 2b, 2c, 3a, 3b and 3c illustrate this relationship graphically using a scatter plot and fitted line of aggregated state-level data on electoral integrity. Figures 2a, 2b and 2c use only 2016 data to show this effect while indicating where each U.S. state is represented on the curve. Figures 3a, 3b and 3c show the effects using both 2016 and 2018 merged data. These graphs clearly document a curvilinear (U-shaped) effect between poverty and electoral integrity using state-level electoral integrity data. Interestingly, we observe that U.S. states such Alabama, North Carolina, and Tennessee are at the lowest point in the relationship between poverty and electoral integrity whereas states such as Kentucky, Mississippi, and New Mexico are examples of states with more extreme levels of poverty that exhibit comparatively higher electoral integrity than states at the lowest point of the curve.

----Insert Figures 2a, 2b, 2c, 3a, 3b, and 3c about here----

----Insert Figure 4a, 4b and 4c about here----

Furthermore, using the margins postestimation command in Stata statistical software, we also plotted the predicted values of the state electoral integrity rating, PEI index of electoral integrity, and fairness of electoral laws. The results are shown with 95% confidence intervals in Figures 4a, 4b and 4c. These results further document the predicted curvilinear relationship between poverty and electoral integrity. All three graphs show that the turning point where the effect of poverty on electoral integrity transforms from negative to positive is around 16 – 17% poverty rate (i.e., around 90 – 95 percentile). Overall, these results highlighted above show strong support for our key hypothesis in this study.

To tease out the specific mechanisms that could mediate the effect of poverty on electoral integrity, we suggested that *fairness of electoral laws* could serve as a substantial mechanism for mitigating the adverse effects of poverty on electoral integrity, especially in the context of an advanced democracy such as the United States. To assess this effect, we follow the procedure outlined by Zhao, Lynch, and Chen (2010), which is a more robust extension of the well-documented Baron and Kenny (1986) method. That is, we first assessed the impact of poverty on the expert perception of *fairness of electoral laws* in the state (i.e., X → M; $\beta = -1.591, p<0.01$), and then assessed the potential influence of *fairness of electoral laws* on the main direct measure of *electoral integrity* in step 2 (i.e., M→Y; *state rating:* $\beta = 0.036, p<0.01$). In the third step (X→Y), we entered both *poverty* and *fairness of electoral laws* in the equation predicting *electoral integrity*. The results in this third step show that poverty has a negative and statistically significant effect on the measure of *state rating on electoral integrity* ($\beta = -0.058, p<0.1$). These results as documented in Appendix A show that the coefficients are in line with our reasoning about the potential for mediation.

The above third step in the Baron and Kenny (1986) is consistent with the first step outlined in the Zhao Lynch, and Chen (2010) method. The results of the Monte Carlo tests confirm that this step is indeed significant. Moreover, since the coefficients point in the same direction, the mediation is partial (i.e., complementary mediation, see Zhao, Lynch, & Chen, 2010). Specifically, *fairness of electoral laws* significantly mediates the effects of poverty on *state rating on electoral integrity*, with approximately 50% of the total effect being mediated. This finding confirms our expectations that the core mechanisms responsible for influence of poverty on electoral integrity are primarily based on the perception of electoral laws in the state.

**4.3 Robustness checks using subsample expert-level multivariate analyses**

Table 3 shows robustness tests for the subset of data based on respondents (i.e., experts) that supported winning versus losing voters. This effort is intended to rule out the alternative explanation of "loser' effect" or "winner effect", whereby supporters of the losing candidate are more likely than supporters of winning candidates to cite problems of electoral integrity to justify their candidate's defeat (Cantu & Garcia-Ponce, 2015; Mongrain, 2023; Sinclair et al., 2018). The results remain very consistent in respect of the negative effect of state poverty on electoral integrity across Models 11-16, with varying degrees of statistical significance ($p<0.01$ – $p<0.1$). This is regardless of whether the expert is supporting the winner or loser of the election. The effect of the squared term of poverty is consistent with our predictions in Models 11, 13, 14, 15 and 16. Surprisingly, the squared term of poverty in relation to *state rating on electoral integrity* is positive but statistically insignificant in Model 12, which is the subsample of the experts that supported the losing candidate. Overall, there are slight differences across the results, but the main findings remain broadly consistent in these subsets of the data.

**4.4 Robustness checks using aggregated state- level, panel data analysis**

In addition to the primary expert-level analysis, we also conducted panel data analysis on the aggregated state-level data of expert responses to questions on electoral integrity. To attend to this panel-data structure, we implemented OLS estimators with panel-corrected standard errors (PCSE; Blackwell, 2005). This approach leverages a generalized least squares (GLS) estimator corrected for first-order serially correlated residuals that are panel specific (see Cumming, Johan, Oberst & Uzuegbunam, 2020). This method allows us to account for state-level specific factors that also affect electoral integrity but are not captured in our equations.

The results of this panel data analysis are reported in Table 4. Consistent with our predictions, the direct effect of poverty rate in a state is negative and statistically significant across Models 18 – 26 ($p<01$ or $p<0.05$). This effect is also economically meaningful. One standard deviation increase in the poverty level is associated with approximately 2.27% - 5.15% decrease in electoral integrity at the state level. The coefficient of the squared term of poverty is also consistent with our predictions. It is positive and statistically significant ($p<01$ or $p<0.05$) in all models in Table 4 except Model 20 and Model 26. Post estimation margin plots also demonstrated the U-shaped relationship between poverty and *State Rating on Electoral Integrity* and *PEI Index on Electoral Integrity*. For brevity, these plots are not included but are available from the authors on request.

We note that the analyses reported in previous Tables (e.g., Table 4) is based on 49 states (not 50 states and District of Columbia, i.e., 51 *states*). The missing observations in our analysis are Washington DC (District of Columbia) and Nebraska. Both Nebraska and the District of Columbia have missing data related to control variables on party information for the legislative and governor positions. First, the District of Columbia observations drop out of the regressions because state legislative and governor party affiliation are not available for the District of Columbia. We note that the District of Columbia has no governor, and is a unicameral legislature, meaning it has only one chamber called the council. Second, the State of Nebraska is also unicameral (that state only has an upper house) and conducts non-partisan state elections for its upper house members. In robustness checks, we re-ran the analysis without the affected controls (and including District of Columbia). We confirm that the results are still consistent. Table 5 reports the findings excluding control variables pertaining to the effect of divided government in state houses and governor party affiliation.

Finally, we checked the sensitivity of the results to fluctuations in turnouts on Presidential election year (i.e., 2016) when compared to midterm election year (i.e., 2018). The findings reported in Table 6 indicate that the results are broadly consistent for the separate samples, though they appear steadier for 2018 midterm elections.

----Insert Tables 4 5, and 6 about here----

We performed additional robustness checks, including checking subsamples based on the number of responses.[1] We did not detect any systematic bias, although with some of the smaller subsamples, the squared term (U-shaped effect) was insignificant in the regressions, likely due to the smaller sample size. Those additional checks are available on request.

---
[1] We owe thanks to a helpful reviewer for this suggestion.

## 5. CONCLUSIONS

The merging of US electoral integrity data with US demographic data yields interesting insights into regional differences in perceived electoral fairness. Our findings extend prior literature that has focused on how poverty might affect electoral integrity in developing countries (Jensen & Justesen, 2014), by theoretically and empirically documenting its curvilinear manifestation across different states in an advanced economy. The conundrum that this study reveals is that states with moderate levels of poverty may be more susceptible to electoral malpractices than states with low or high levels of poverty because the internal and external demand for electoral integrity is too low in these states, while the cost of rendering electoral integrity is substantially higher. Ironically, poor electoral integrity ultimately diminishes the economic viability of these states, which might eventually lead to more attention to their electoral practices. Gaining awareness of this conundrum should spur policy makers and external organizations toward directing much needed electoral attention and resources to these states. The evidence here reveals mechanisms that have helped in the past, including the strengthening of electoral laws. For example, this research provides strong empirical evidence in support of recent proposals for more uniformity and stringent standards in laws governing how state and local officials select voting machines (Weichelt, 2022). Such strengthening of the legal standards for elections across states and municipalities is likely to increase perceptions of electoral integrity, regardless of poverty conditions in the geographies.

This study is not without limitations. We recognize that elections are often run at the local level. However, our electoral integrity data is currently limited at the state level. This presents opportunities for future research to examine the relationship between poverty and electoral integrity at the micro level. By delving into this level of analysis, future research might also be able to uncover how poverty interacts with other local customs and cultures to influence electoral integrity.

**FIGURE 1.** Theoretical predictions regarding poverty rates and electoral integrity

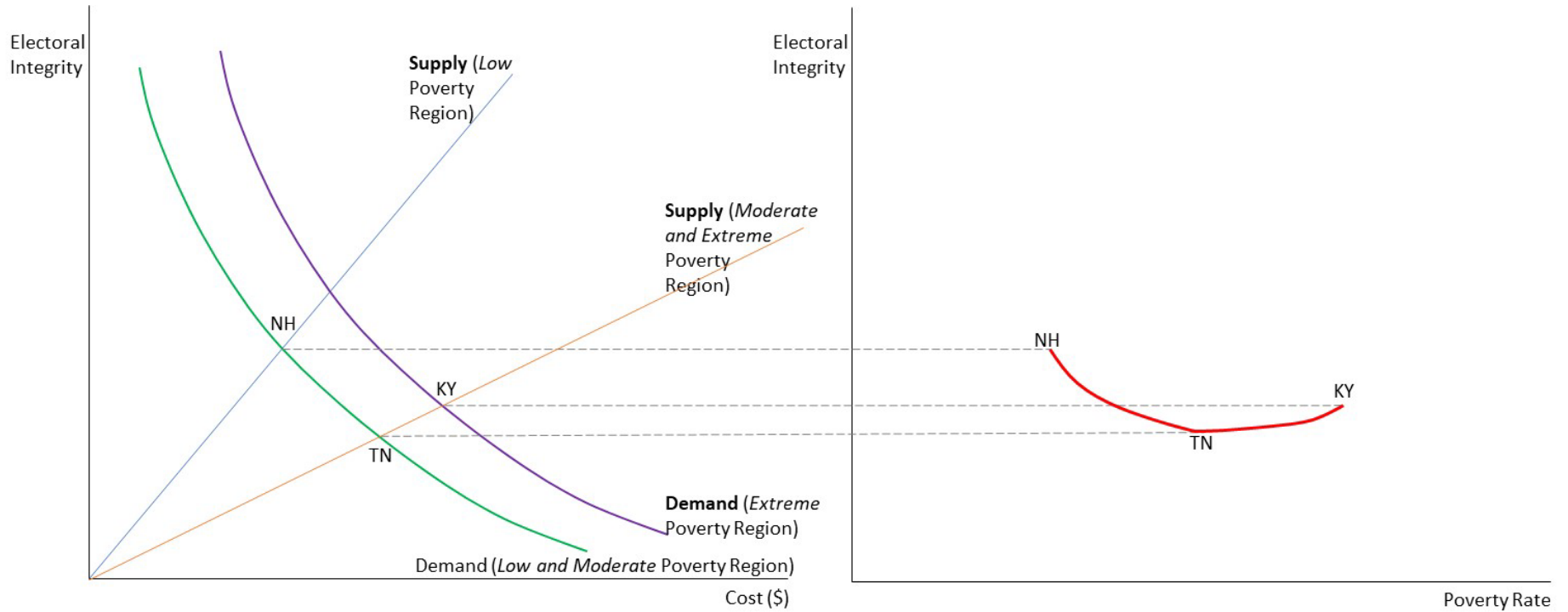

Note: The left-hand-side figure shows the supply of electoral integrity and demand for electoral integrity as a function of costs. Demand for electoral integrity slopes downward as taxpayers are willing to accept less integrity if it costs their tax dollars more. Demand is convex because citizens will prefer to have more electoral integrity despite the costs in cases of very low levels of electoral integrity; conversely, if it is very cheap to provide, citizens prefer much higher levels of electoral integrity. Supply slopes upwards because the supply of more electoral integrity costs more. Supply of electoral integrity costs more for any given amount of electoral integrity in moderate and extreme poverty regions because those regions do not have the same infrastructure and monitoring capacity to provide electoral integrity at a cheaper cost. Demand for electoral integrity is higher in extreme poverty regions that attract external scrutiny from election monitoring agencies. NH (New Hampshire), TN (Tennessee), and KY (Kentucky) are indicated as examples of states with low, moderate, and extreme poverty rates, respectively. The resulting right-hand-side figure is the predicted relation between poverty rates in a region and electoral integrity.

**FIGURE 2a:** Plot of State-Level Electoral Integrity Rating on State-Level Poverty (2016 only)

**FIGURE 2b:** Plot of State-Level PEI Index for Electoral Integrity on State-Level Poverty (2016 only)

**FIGURE 2c:** Plot of State-Level Electoral Laws Index on State-Level Poverty (2016 only)

**FIGURE 3a:** Plot of State-Level Electoral Integrity Rating on State-Level Poverty (2016 & 2018)

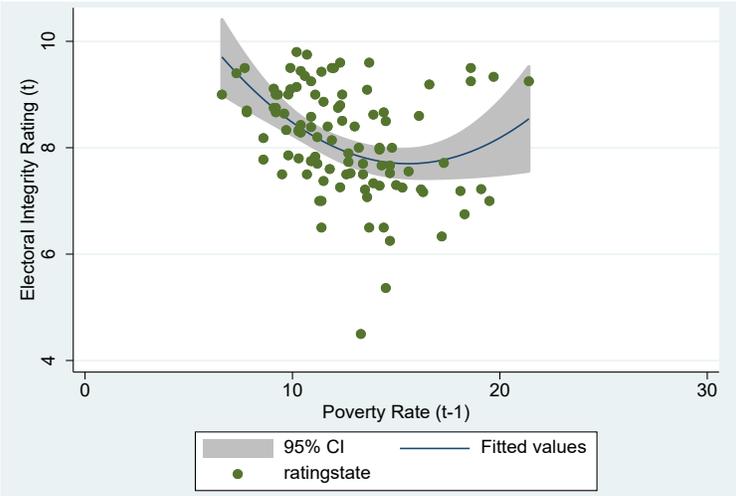

**FIGURE 3b:** Plot of State-Level PEI Index on State-Level Poverty (2016 & 2018)

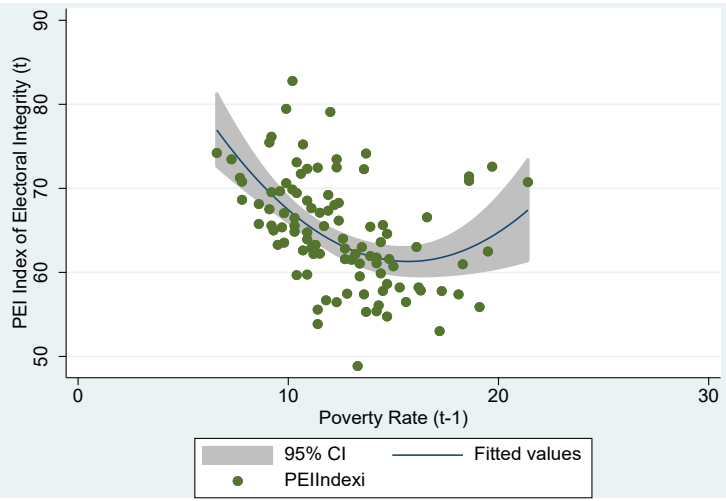

**FIGURE 3c:** Plot of State-Level Electoral Laws Index on State-Level Poverty (2016 & 2018)

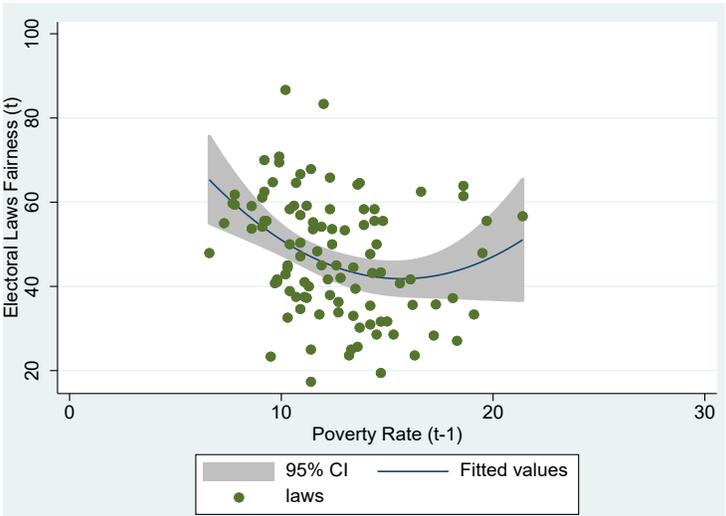

**FIGURE 4a:** Predicted Impact of Poverty on State-level Electoral Integrity (Expert-level data)

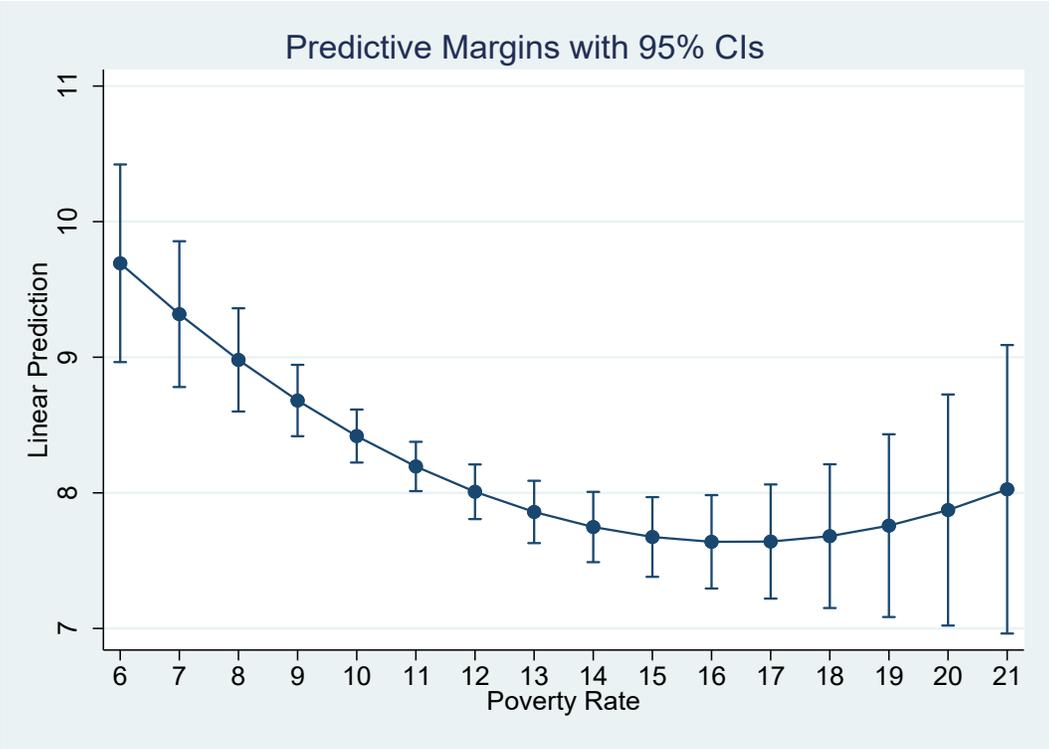

**FIGURE 4b:** Predicted Impact of Poverty on PEI Index of Electoral Integrity (Expert-level data)

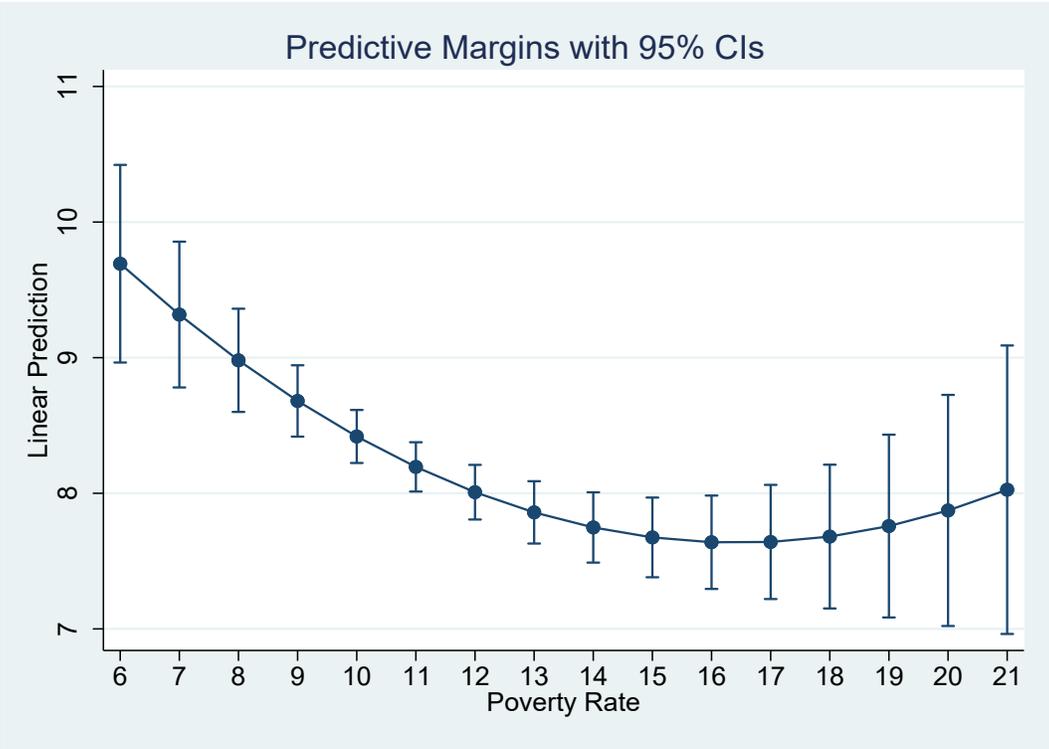

**FIGURE 4c:** Predicted Impact of Poverty on Electoral Law Index (Expert-level data)

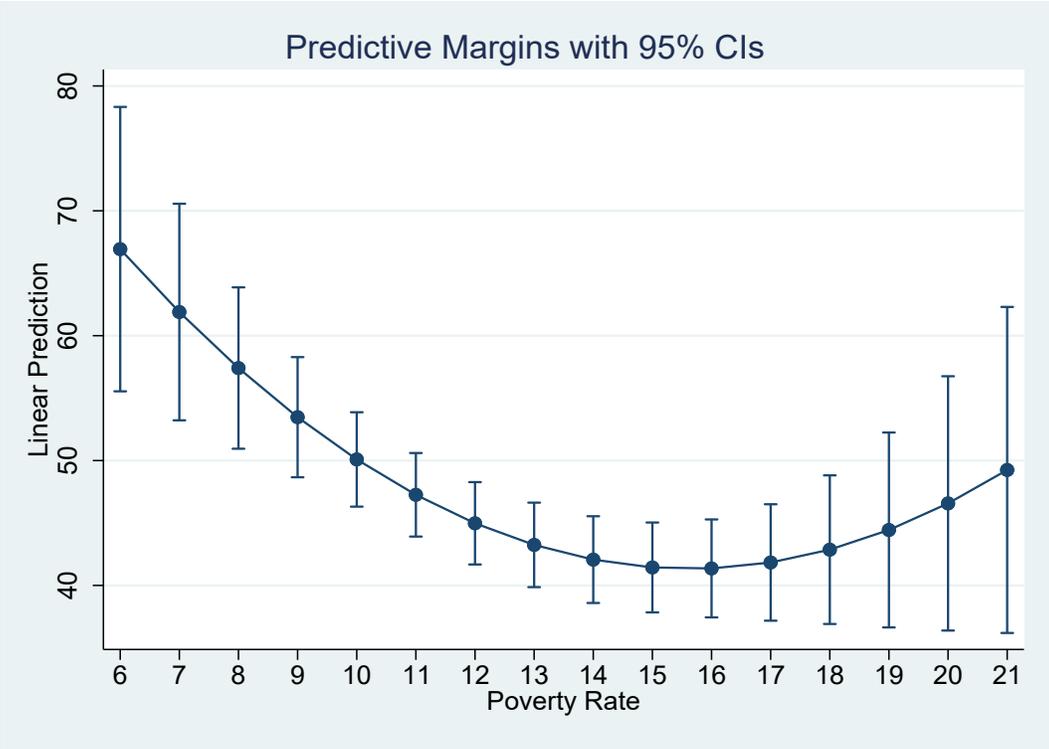

**TABLE 1**: Summary Statistics and Bivariate Correlations of Study Variables

| | 1 | 2 | 3 | 4 | 5 | 6 | 7 | 8 | 9 | 10 | 11 | 12 | 13 | 14 |
|---|---|---|---|---|---|---|---|---|---|---|---|---|---|---|
| 1. State rating | | | | | | | | | | | | | | |
| 2. PEI index-electoral integrity | 0.721 | | | | | | | | | | | | | |
| 3. Electoral laws index | 0.515 | 0.649 | | | | | | | | | | | | |
| 4. State poverty rate | -0.164 | -0.218 | -0.151 | | | | | | | | | | | |
| 5. Governor is Democrat | 0.074 | 0.080 | 0.109 | -0.134 | | | | | | | | | | |
| 6. State house/senate Dems | 0.269 | 0.235 | 0.271 | -0.206 | 0.368 | | | | | | | | | |
| 7. Country electoral integrity | 0.450 | 0.390 | 0.271 | 0.077 | -0.101 | -0.108 | | | | | | | | |
| 8. No rigged elections | -0.020 | 0.068 | -0.024 | -0.183 | -0.024 | -0.011 | -0.200 | | | | | | | |
| 9. Expert's gender (F=1) | -0.093 | -0.173 | -0.108 | 0.030 | 0.076 | 0.049 | -0.168 | 0.007 | | | | | | |
| 10. Expert is US Citizen | 0.032 | 0.054 | 0.037 | -0.034 | 0.029 | 0.034 | 0.006 | 0.014 | -0.046 | | | | | |
| 11. Expert was born in state | 0.005 | -0.041 | -0.017 | -0.013 | 0.058 | 0.102 | -0.037 | 0.014 | -0.021 | 0.070 | | | | |
| 12. Expert lived in state | -0.059 | -0.033 | -0.046 | -0.042 | 0.062 | 0.118 | -0.113 | 0.046 | -0.051 | 0.144 | 0.328 | | | |
| 13. Supported winner | 0.004 | 0.102 | 0.009 | -0.176 | 0.036 | 0.011 | -0.106 | 0.639 | 0.007 | 0.110 | 0.009 | 0.075 | | |
| 14. Supported loser | 0.024 | -0.048 | 0.019 | 0.162 | 0.010 | 0.033 | 0.073 | -0.609 | 0.053 | 0.149 | 0.002 | -0.024 | -0.646 | |
| Mean | 8.018 | 63.552 | 45.668 | 12.708 | 0.414 | 0.944 | 6.361 | 2.858 | 0.303 | 0.975 | 0.165 | 3.717 | 0.323 | 0.467 |
| S.D. | 1.815 | 11.202 | 25.475 | 2.439 | 0.493 | 0.347 | 2.108 | 1.655 | 0.46 | 0.155 | 0.371 | 0.758 | 0.468 | 0.499 |
| Min | 1 | 15.306 | 0 | 6.6 | 0 | 0.25 | 1 | 1 | 0 | 0 | 0 | 0 | 0 | 0 |
| Max | 10 | 94.311 | 100 | 21.4 | 1 | 1.9 | 10 | 5 | 1 | 1 | 1 | 5 | 1 | 1 |

Correlations ≥ |0.09|, |0.06| and |0.05| are significant at the 1%, 5%, and 10% levels respectively.

TABLE 2: Multivariate Regression Results of State-Level Poverty, and Experts Rating of Electoral Integrity

|  | Model 1 | Model 2 | Model 3 | Model 4 | Model 5 | Model 6 | Model 7 |
|---|---|---|---|---|---|---|---|
|  | DV: State Electoral Integrity Rating | | | | DV: PEIIndexi of Electoral Integrity | | |
| Electoral laws |  |  |  | 0.0229*** |  |  |  |
|  |  |  |  | [0.0024] |  |  |  |
| State poverty rate |  | -0.1255*** | -0.1410*** | -0.0966*** |  | -0.8641*** | -1.0102*** |
|  |  | [0.0397] | [0.0352] | [0.0265] |  | [0.2465] | [0.1836] |
| Poverty rate squared |  |  | 0.0189** | 0.0126** |  |  | 0.1595*** |
|  |  |  | [0.0074] | [0.0056] |  |  | [0.0417] |
| Governor is Democrat | 0.0806 | 0.0290 | 0.0445 | 0.0015 | 0.8444 | 0.5153 | 0.6051 |
|  | [0.1857] | [0.1795] | [0.1794] | [0.1384] | [1.2796] | [1.2263] | [1.1790] |
| State house/senate Democrats | 1.6557*** | 1.4936*** | 1.5057*** | 1.0653*** | 9.2742*** | 8.1380*** | 8.2330*** |
|  | [0.2505] | [0.2174] | [0.2185] | [0.1428] | [1.4746] | [1.4520] | [1.4405] |
| Country electoral integrity rating | 0.4266*** | 0.4304*** | 0.4270*** | 0.3503*** | 2.2821*** | 2.3088*** | 2.2737*** |
|  | [0.0377] | [0.0390] | [0.0390] | [0.0411] | [0.1814] | [0.1829] | [0.1805] |
| No rigged elections | 0.1657* | 0.1647* | 0.1547* | 0.1396* | 0.4765 | 0.4709 | 0.3879 |
|  | [0.0979] | [0.0935] | [0.0919] | [0.0739] | [0.4112] | [0.3916] | [0.3779] |
| Expert's sex (Female=1) | -0.0979 | -0.0692 | -0.0655 | 0.0647 | -2.5594*** | -2.3873*** | -2.3595*** |
|  | [0.0896] | [0.0869] | [0.0870] | [0.0792] | [0.5635] | [0.5532] | [0.5509] |
| Expert is US citizen (US citizen=1) | 0.1037 | 0.0495 | 0.0332 | -0.0176 | 1.1501 | 0.8062 | 0.6005 |
|  | [0.2458] | [0.2506] | [0.2496] | [0.2140] | [2.5396] | [2.5278] | [2.5407] |
| Expert was born in state | 0.0325 | 0.0471 | 0.0786 | 0.0844 | -1.5824* | -1.4924* | -1.2089 |
|  | [0.1448] | [0.1400] | [0.1440] | [0.1174] | [0.8670] | [0.8511] | [0.8716] |
| Expert lived in state | -0.0811 | -0.0726 | -0.0649 | -0.0431 | -0.3223 | -0.2503 | -0.1880 |
|  | [0.0814] | [0.0753] | [0.0761] | [0.0740] | [0.4850] | [0.4535] | [0.4591] |
| Supported | -0.0426 | -0.0450 | -0.0481 | -0.0416 | -0.4303* | -0.4382* | -0.4838* |
|  | [0.0395] | [0.0378] | [0.0376] | [0.0325] | [0.2534] | [0.2431] | [0.2424] |
| Year is 2018 | -0.3296 | -0.4534 | -0.4372 | -0.4180 | 2.2042 | 1.3541 | 1.4267 |
|  | [0.3562] | [0.3658] | [0.3559] | [0.2961] | [1.3984] | [1.4893] | [1.3892] |
| Constant | 3.7008*** | 3.9164*** | 3.8192*** | 3.7042*** | 39.8000*** | 41.2092*** | 40.5859*** |
|  | [0.7249] | [0.6905] | [0.6826] | [0.6158] | [4.2636] | [4.0965] | [4.0018] |
| Observations (States) | 1,208 (49) | 1,208 (49) | 1,208 (49) | 1,152 (49) | 1,202 (49) | 1,202 (49) | 1,202 (49) |
| Adjusted R-squared | 0.310 | 0.335 | 0.344 | 0.429 | 0.278 | 0.309 | 0.326 |

The table reports parameter coefficient estimates. Standard errors (clustered by state) are in parentheses. ***p<0.01, ** p<0.05 and * p<0.1 indicate significance, using two-tailed tests. The dependent variable is assessed at the expert-level in year, t+1, and the independent variables are assessed in year, t (i.e., year is 2015 or 2017 respectively).

**TABLE 2 Cont'd**: Multivariate Regression Results

|  | Model 8 | Model 9 | Model 10 |
|---|---|---|---|
|  | | DV: Electoral Laws Index | |
| State poverty rate |  | **-1.3445**\*\* | **-1.6155**\*\*\* |
|  |  | [0.5649] | [0.4860] |
| Poverty rate squared |  |  | **0.2756**\*\* |
|  |  |  | [0.1046] |
| Governor is Democrat | 2.4787 | 1.9335 | 2.0322 |
|  | [2.6909] | [2.7569] | [2.6841] |
| State house/senate Democrats | 22.0388*** | 20.3244*** | 20.5295*** |
|  | [3.6257] | [3.6898] | [3.5476] |
| Country electoral integrity rating | 3.4888*** | 3.5334*** | 3.4642*** |
|  | [0.3641] | [0.3542] | [0.3495] |
| No rigged elections | -0.4253 | -0.4254 | -0.5639 |
|  | [0.9437] | [0.9138] | [0.9125] |
| Expert's sex (Female=1) | -4.5176*** | -4.2551*** | -4.2075*** |
|  | [1.4582] | [1.4539] | [1.4768] |
| Expert is US citizen (US citizen=1) | 1.6297 | 1.1726 | 0.9230 |
|  | [3.6540] | [3.5583] | [3.5480] |
| Expert was born in state | -1.8309 | -1.6606 | -1.1733 |
|  | [2.1224] | [2.1443] | [2.1792] |
| Expert lived in state | -0.8972 | -0.8580 | -0.7599 |
|  | [1.1256] | [1.1001] | [1.1306] |
| Supported | -0.3575 | -0.3799 | -0.4412 |
|  | [0.4727] | [0.4600] | [0.4449] |
| Year is 2018 | 4.1351 | 2.8401 | 2.9677 |
|  | [2.5056] | [2.7171] | [2.6468] |
| Constant | 4.9115 | 7.2272 | 6.0659 |
|  | [7.7757] | [7.3063] | [7.2733] |
| State fixed effects | No | No | No |
| Observations (States) | 1,161 (49) | 1,161 (49) | 1,161 (49) |
| Adjusted R-squared | 0.169 | 0.183 | 0.193 |

The table reports parameter coefficient estimates. Standard errors (clustered by state) are in parentheses. ***p<0.01, **p<0.05 and * p<0.1 indicate significance, using two-tailed tests. The dependent variable is assessed at the expert-level in year, t+1, and the independent variables are assessed in year, t (i.e., year is 2015 or 2017 respectively).

**TABLE 3:** Subsample Analysis Comparing Winning vs. Losing Supporters

|  | Model 11 | Model 12 | Model 13 | Model 14 | Model 15 | Model 16 |
|---|---|---|---|---|---|---|
|  | Electoral Integrity Rating | | PEIIndexi of Electoral Integrity | | Electoral Laws Index | |
| Subsample→ | Supported winner | Supported loser | Supported winner | Supported loser | Supported winner | Supported loser |
| State poverty rate | **-0.1057*** | **-0.0978*** | **-0.8797*** | **-0.7720*** | **-1.4326**  | **-0.9949*** |
|  | [0.0370] | [0.0299] | [0.2229] | [0.2154] | [0.5645] | [0.5786] |
| Poverty rate squared | **0.0189**  | 0.0068 | **0.1293**  | **0.1246**  | **0.3696**  | **0.2450*** |
|  | [0.0082] | [0.0069] | [0.0532] | [0.0545] | [0.1541] | [0.1250] |
| Governor is Democrat | -0.0673 | 0.3222* | 0.2524 | 1.5402 | -0.9233 | 4.6246 |
|  | [0.2469] | [0.1763] | [1.2923] | [1.4298] | [3.1342] | [3.0748] |
| State house/senate Democrats | 1.3706*** | 1.3085*** | 7.2386*** | 7.9697*** | 22.3973*** | 22.3768*** |
|  | [0.2266] | [0.2373] | [1.3143] | [1.8108] | [3.9465] | [3.8129] |
| Country electoral integrity rating | 0.2071*** | 0.3664*** | 1.2632*** | 1.7199*** | 2.4201*** | 2.3246*** |
|  | [0.0447] | [0.0514] | [0.2319] | [0.2621] | [0.5869] | [0.5487] |
| No rigged elections | 0.8593*** | -0.3348*** | 4.3852*** | -2.0920*** | 5.2938*** | -4.9720*** |
|  | [0.1527] | [0.1093] | [0.6809] | [0.7101] | [1.2117] | [1.5344] |
| Expert's sex (Female=1) | 0.1545 | -0.2325** | -0.2330 | -4.3182*** | -4.0883* | -4.4737** |
|  | [0.1209] | [0.1140] | [0.8542] | [0.7210] | [2.0433] | [2.1501] |
| Expert was born in state | -0.1713 | 0.1016 | -0.4457 | -0.9949 | -1.6698 | -0.1117 |
|  | [0.2307] | [0.1624] | [1.1348] | [1.1027] | [3.0187] | [2.3979] |
| Expert lived in state | -0.0832 | 0.0016 | -1.0159 | 0.2165 | -3.8971** | 0.7311 |
|  | [0.1167] | [0.0967] | [0.7013] | [0.6169] | [1.8742] | [1.6773] |
| Year is 2018 | -2.6597*** | 1.3359*** | -12.5663*** | 12.2437*** | -24.1808*** | 33.3123*** |
|  | [0.5780] | [0.4459] | [2.7982] | [3.6915] | [6.8422] | [4.7430] |
| Constant | 4.4855*** | 4.8432*** | 46.6735*** | 46.8310*** | 24.4584** | 11.7582 |
|  | [0.6421] | [0.7302] | [4.3487] | [3.5768] | [11.6769] | [10.0665] |
| Observations (States) | 400 (49) | 558 (48) | 400 (49) | 562 (48) | 376 (49) | 540 (48) |
| Adjusted R-squared | 0.435 | 0.425 | 0.406 | 0.393 | 0.252 | 0.255 |

The table reports parameter coefficient estimates. Standard errors (clustered by state) are in parentheses. ***p<0.01, ** p<0.05 and * p<0.1 indicate significance, using two-tailed tests. The dependent variable is assessed at the expert-level in year, t+1, and the independent variables are assessed in year, t (i.e., year is 2015 or 2017 respectively).

**TABLE 4:** State-level Panel Corrected Standard Error (PCSE) Estimates of Poverty Impact on Electoral Integrity

|  | Model 17 | Model 18 | Model 19 | Model 20 | Model 21 | Model 22 | Model 23 | Model 24 | Model 25 | Model 26 |
|---|---|---|---|---|---|---|---|---|---|---|
|  | DV: State Electoral Integrity Rating | | | | DV: PEIIndexi of Electoral Integrity | | | DV: Electoral Laws Index | | |
| Electoral laws |  |  |  | 0.0338*** |  |  |  |  |  |  |
|  |  |  |  | [0.0031] |  |  |  |  |  |  |
| State poverty rate |  | -0.0617*** | -0.0859*** | -0.0467** |  | -0.5445*** | -0.8283*** |  | -0.8114*** | -1.1577*** |
|  |  | [0.0188] | [0.0103] | [0.0188] |  | [0.1187] | [0.0755] |  | [0.1285] | [0.2254] |
| Poverty rate squared |  |  | 0.0070*** | 0.0036 |  |  | 0.0824** |  |  | 0.1005 |
|  |  |  | [0.0027] | [0.0039] |  |  | [0.0348] |  |  | [0.0887] |
| Governor is Democrat | 0.3705*** | 0.2881*** | 0.2494** | 0.1872** | 2.6669*** | 1.9402*** | 1.4861** | 3.4745* | 2.3915 | 1.8374 |
|  | [0.1274] | [0.1063] | [0.1177] | [0.0819] | [0.7698] | [0.6787] | [0.7446] | [1.7812] | [1.5539] | [1.5630] |
| State house/senate Democrats | 0.7043*** | 0.6885*** | 0.6781*** | 0.2745** | 4.0487*** | 3.9095*** | 3.7879*** | 12.2798*** | 12.0725*** | 11.9241*** |
|  | [0.1366] | [0.1383] | [0.1382] | [0.1133] | [0.4317] | [0.2911] | [0.4177] | [3.1890] | [2.9549] | [3.0044] |
| Country electoral integrity rating | 0.1758* | 0.2088** | 0.1840** | 0.0945 | 1.2297** | 1.5204*** | 1.2294*** | 2.5660 | 2.9992** | 2.6441* |
|  | [0.1032] | [0.0913] | [0.0817] | [0.1079] | [0.5972] | [0.4904] | [0.3974] | [1.5920] | [1.3870] | [1.4362] |
| No rigged elections | 1.3688*** | 1.3022*** | 1.2594*** | 0.9509*** | 8.0124*** | 7.4250*** | 6.9226*** | 10.6040*** | 9.7286*** | 9.1155*** |
|  | [0.2183] | [0.1970] | [0.1925] | [0.1076] | [1.6111] | [1.3987] | [1.3090] | [3.4431] | [3.1788] | [2.8606] |
| Expert is US citizen (US citizen=1) | 1.3186 | 0.8963 | 0.8399 | -0.2186 | 11.8739 | 8.1501 | 7.4884 | 37.6300** | 32.0807** | 31.2732** |
|  | [1.3805] | [1.2207] | [1.1786] | [0.9239] | [9.2669] | [7.8448] | [6.8293] | [15.6890] | [13.4799] | [12.3329] |
| Expert was born in state | -0.1467 | -0.1197 | -0.0105 | 0.1104 | -3.6948** | -3.4560* | -2.1748 | -5.4898 | -5.1340 | -3.5706 |
|  | [0.1659] | [0.1699] | [0.2634] | [0.2167] | [1.8637] | [1.7779] | [2.3613] | [7.9814] | [7.9742] | [8.6210] |
| Expert lived in state | -0.1799 | -0.1952 | -0.2047 | -0.0596 | -1.5925*** | -1.7267*** | -1.8389** | -3.9499* | -4.1500* | -4.2869* |
|  | [0.1368] | [0.1327] | [0.1368] | [0.1730] | [0.4402] | [0.4740] | [0.7316] | [2.1633] | [2.1256] | [2.4067] |
| Supported | 0.2867** | 0.2551*** | 0.2307** | 0.0922 | 1.8426*** | 1.5646*** | 1.2782*** | 4.8576*** | 4.4434*** | 4.0938*** |
|  | [0.1284] | [0.0923] | [0.0949] | [0.0742] | [0.7117] | [0.4179] | [0.3113] | [1.4156] | [1.0141] | [0.8138] |
| Year is 2018 | 0.2660* | 0.2105 | 0.1656 | -0.0208 | 4.3556*** | 3.8660*** | 3.3395*** | 6.8784*** | 6.1488*** | 5.5063*** |
|  | [0.1490] | [0.1341] | [0.1238] | [0.1139] | [0.7204] | [0.5142] | [0.3151] | [1.5696] | [1.1864] | [1.0123] |
| Constant | -1.2252 | -0.4970 | -0.0210 | 1.6248 | 5.1345 | 11.5570 | 17.1403 | -65.0094* | -55.4383* | -48.6253* |
|  | [2.8491] | [2.5774] | [2.5004] | [1.7850] | [16.0312] | [13.2017] | [11.0545] | [36.2467] | [33.0197] | [27.8726] |
| Observations (states) | 98 (49) | 98 (49) | 98 (49) | 98 (49) | 98 (49) | 98 (49) | 98 (49) | 98 (49) | 98 (49) | 98 (49) |
| R-squared | 0.561 | 0.590 | 0.595 | 0.741 | 0.514 | 0.568 | 0.586 | 0.330 | 0.356 | 0.361 |

The table reports parameter coefficient estimates. Standard errors (clustered by state) are in parentheses. ***p<0.01, ** p<0.05 and * p<0.1 indicate significance, using two-tailed tests. The dependent variable is assessed at the state-level in year, t+1, and the independent variables are assessed at the state-level in year, t.

**TABLE 5:** State-level Panel Corrected Standard Error (PCSE) Estimates of Poverty Impact on Electoral Integrity (Excluding State House and Governor controls)

|  | Model 27 State Electoral Integrity Rating | Model 28 PEIIndexi | Model 29 Electoral Laws Index |
|---|---|---|---|
| State poverty rate | -0.1069*** | -0.9460*** | -1.3453*** |
|  | [0.0142] | [0.1253] | [0.4024] |
| Poverty rate squared | 0.0102** | 0.1013** | 0.1355 |
|  | [0.0041] | [0.0475] | [0.1111] |
| Country electoral integrity rating | 0.0190 | 0.1373 | 0.1366 |
|  | [0.0256] | [0.5045] | [1.9371] |
| No rigged elections | 1.4521*** | 8.0290*** | 12.8490*** |
|  | [0.2076] | [1.4980] | [3.3522] |
| Expert is US citizen (US citizen=1) | 1.5357 | 11.8782 | 40.0564*** |
|  | [1.7395] | [9.5994] | [15.3061] |
| Expert was born in state | -0.2723 | -3.4466 | -6.9558 |
|  | [0.3193] | [3.0534] | [10.2564] |
| Expert lived in state | -0.1336 | -1.3846 | -2.8625 |
|  | [0.1284] | [0.9673] | [3.0530] |
| Supported | 0.1435 | 0.9557** | 2.9875*** |
|  | [0.0972] | [0.4316] | [0.8594] |
| Year is 2018 | -0.0157 | 2.2000*** | 2.7376*** |
|  | [0.0789] | [0.2820] | [0.6781] |
| Constant | 0.2676 | 18.6133 | -47.3161 |
|  | [2.9638] | [14.2642] | [33.2280] |
| Observations | 102 | 102 | 102 |
| R-squared | 0.506 | 0.514 | 0.275 |
| Number of states | 51 | 51 | 51 |

The table reports parameter coefficient estimates. Standard errors (clustered by state) are in parentheses. ***p<0.01, ** p<0.05 and * p<0.1 indicate significance, using two-tailed tests. The dependent variable is assessed at the state-level in year, t+1, and the independent variables are assessed at the state-level in year, t.

**TABLE 6:** Robustness Checks using 2016 Presidential and 2018 midterm elections subsamples

|  | Model 30 State Electoral Integrity Rating | Model 31 PEIIndexi | Model 32 Electoral Laws Index | Model 33 State Electoral Integrity Rating | Model 34 PEIIndexi | Model 35 Electoral Laws Index |
|---|---|---|---|---|---|---|
|  | 2016 only | 2016 only | 2016 only | 2018 only | 2018 only | 2018 only |
| State poverty rate | -0.0982*** | -0.7235*** | -1.2379** | -0.1168*** | -0.9682*** | -1.4502** |
|  | [0.0257] | [0.1790] | [0.4970] | [0.0318] | [0.2023] | [0.5552] |
| Poverty rate squared | 0.0060 | 0.1124** | 0.2278* | 0.0208*** | 0.1380*** | 0.2593** |
|  | [0.0075] | [0.0550] | [0.1233] | [0.0069] | [0.0428] | [0.1169] |
| Governor is Democrat | 0.2375 | 1.2849 | 4.4784 | -0.1804 | -0.2178 | -1.1696 |
|  | [0.1777] | [1.3808] | [2.8052] | [0.2103] | [1.2452] | [3.2079] |
| State house/senate Democrats | 1.2445*** | 6.3674*** | 19.3061*** | 1.2036*** | 6.4392*** | 17.1109*** |
|  | [0.2193] | [1.7352] | [3.6869] | [0.1886] | [1.2804] | [4.1007] |
| Country electoral integrity rating | 0.3872*** | 1.6698*** | 2.5263*** | 0.2514*** | 1.5870*** | 2.5165*** |
|  | [0.0453] | [0.1969] | [0.4688] | [0.0557] | [0.2676] | [0.5445] |
| No rigged elections | -0.3990*** | -3.5365*** | -6.2344*** | 0.8843*** | 4.8519*** | 5.9157*** |
|  | [0.0986] | [0.6249] | [1.1481] | [0.1377] | [0.6727] | [1.2467] |
| Expert's sex (Female=1) | -0.1491 | -3.0896*** | -2.6702 | 0.1391 | -0.7415 | -5.3023** |
|  | [0.1021] | [0.7532] | [1.9508] | [0.1250] | [0.8998] | [2.2358] |
| Expert is US citizen (US citizen=1) | -0.2496 | 0.0924 | -0.9930 | 0.7219 | 2.2818 | 5.5157 |
|  | [0.2057] | [2.3330] | [4.3254] | [0.5533] | [3.0508] | [5.6521] |
| Expert was born in state | 0.1551 | -1.5364 | -0.2199 | -0.1438 | -1.3550 | -3.2624 |
|  | [0.1517] | [0.9966] | [2.3512] | [0.1705] | [0.9675] | [2.8840] |
| Expert lived in state | 0.0025 | 0.5155 | 0.4002 | -0.0906 | -0.6742 | -1.2911 |
|  | [0.0860] | [0.5486] | [1.6868] | [0.1153] | [0.6108] | [1.5884] |
| Supported | -0.0433 | -0.4651** | -0.3888 | -0.0237 | -0.2855 | -0.1756 |
|  | [0.0289] | [0.2210] | [0.5355] | [0.0493] | [0.2906] | [0.6840] |
| Constant | 5.2305*** | 50.7288*** | 18.4337** | 0.9665 | 28.0588*** | -11.5727 |
|  | [0.6750] | [4.4236] | [8.9510] | [1.0806] | [5.5777] | [11.1155] |
| Observations | 669 (49) | 666 (49) | 650 (49) | 539 (49) | 536 (49) | 511 (49) |
| Adjusted R-squared | 0.450 | 0.405 | 0.246 | 0.440 | 0.438 | 0.208 |

The table reports parameter coefficient estimates. Standard errors (clustered by state) are in parentheses. ***p<0.01, ** p<0.05 and * p<0.1 indicate significance, using two-tailed tests. The dependent variable is assessed at the state-level in year, t+1, and the independent variables are assessed at the state-level in year, t.

Appendix A: Mediation Testing in the Baron and Kenny (1986) Steps

|  | (1) State rating | (2) Electoral laws | (3) / |
|---|---|---|---|
| Electoral laws | 0.0357*** [0.0018] |  |  |
| State poverty rate | -0.0577*** [0.0183] | -1.5913*** [0.2942] |  |
| Var (e. State rating) |  |  | 2.3847*** [0.0964] |
| Var (e. Electoral laws) |  |  | 632.4560*** [25.5655] |
| Var (e. PEI Index) |  |  |  |
| Constant | 7.1316*** [0.2608] | 65.9665*** [3.8057] |  |
| Observations | 1,224 | 1,224 | 1,224 |

Appendix B: Distribution of Survey Respondents By State and Year

|  | Observations by year | |  |
|---|---|---|---|
| State | 2016 | 2018 | Total |
| AK | 3 | 3 | 6 |
| AL | 6 | 10 | 16 |
| AR | 5 | 4 | 9 |
| AZ | 6 | 7 | 13 |
| CA | 78 | 66 | 144 |
| CO | 10 | 6 | 16 |
| CT | 12 | 12 | 24 |
| DC | 22 | 11 | 33 |
| DE | 3 | 4 | 7 |
| FL | 23 | 16 | 39 |
| GA | 16 | 14 | 30 |
| HI | 4 | 6 | 10 |
| IA | 9 | 9 | 18 |
| ID | 5 | 5 | 10 |
| IL | 36 | 24 | 60 |
| IN | 15 | 16 | 31 |
| KS | 4 | 6 | 10 |
| KY | 4 | 2 | 6 |
| LA | 8 | 5 | 13 |
| MA | 37 | 20 | 57 |
| MD | 14 | 10 | 24 |

| State | | | |
|---|---:|---:|---:|
| ME | 10 | 14 | 24 |
| MI | 24 | 15 | 39 |
| MN | 12 | 11 | 23 |
| MO | 14 | 13 | 27 |
| MS | 9 | 4 | 13 |
| MT | 7 | 9 | 16 |
| NC | 29 | 22 | 51 |
| ND | 2 | 5 | 7 |
| NE | 5 | 7 | 12 |
| NH | 5 | 5 | 10 |
| NJ | 10 | 9 | 19 |
| NM | 3 | 6 | 9 |
| NV | 5 | 5 | 10 |
| NY | 58 | 40 | 98 |
| OH | 28 | 19 | 47 |
| OK | 7 | 6 | 13 |
| OR | 6 | 7 | 13 |
| PA | 39 | 20 | 59 |
| RI | 5 | 4 | 9 |
| SC | 12 | 9 | 21 |
| SD | 3 | 7 | 10 |
| TN | 4 | 8 | 12 |
| TX | 46 | 27 | 73 |
| UT | 3 | 11 | 14 |
| VA | 28 | 13 | 41 |
| VT | 4 | 5 | 9 |
| WA | 7 | 2 | 9 |
| WI | 13 | 6 | 19 |
| WV | 2 | 7 | 9 |
| WY | 6 | 2 | 8 |
| Total | 726 | 574 | 1,300 |